\documentclass[iop]{emulateapj}

\begin{document}
\shorttitle{Cosmic Battery} \shortauthors{Contopoulos et al.}

\def\gsim{\mathrel{\raise.5ex\hbox{$>$}\mkern-14mu
             \lower_{\rm 0}.6ex\hbox{$\sim$}}}
+\def\lsim{\mathrel{\raise.3ex\hbox{$<$}\mkern-14mu
             \lower_{\rm 0}.6ex\hbox{$\sim$}}}

\author{Ioannis Contopoulos\altaffilmark{1,}\altaffilmark{*},
Antonios Nathanail\altaffilmark{1,}\altaffilmark{2} and Matthaios
Katsanikas\altaffilmark{1}}
% and Leela E.
%Koutsantoniou\altaffilmark{1,}\altaffilmark{2}}
\affil{\altaffilmark{1} Research Center for Astronomy and Applied
Mathematics, Academy of Athens, Athens
11527, Greece \\ and \\
\altaffilmark{2}Section of Astrophysics, Astronomy and Mechanics,
Department of Physics, University of Athens, \\
Panepistimiopolis Zografos, Athens 15783, Greece}
%\altaffilmark{2}}
%\affil{NASA/GSFC, Code 663, Greenbelt, MD 20771, USA;
%konstantinos.kalapotharakos, demos.kazanas@nasa.gov}
\altaffiltext{*}{icontop@academyofathens.gr}

\title{The Cosmic Battery in Astrophysical Accretion Disks}

%\altaffiltext{1}{icontop@academyofathens.gr}
%\altaffiltext{2}{}
%\altaffiltext{3}{}

%% Mark off your abstract in the ``abstract'' environment. In the manuscript
%% style, abstract will output a Received/Accepted line after the
%% title and affiliation information. No date will appear since the author
%% does not have this information. The dates will be filled in by the
%% editorial office after submission.

\begin{abstract}
The aberrated radiation pressure at the inner edge of the
accretion disk around an astrophysical black hole imparts a
relative azimuthal velocity on the electrons with respect to the
ions which gives rise to a ring electric current that generates
large scale poloidal magnetic field loops. This is the Cosmic
Battery established by Contopoulos and Kazanas in 1998. In the
present work we perform realistic numerical simulations of this
important astrophysical mechanism in advection-dominated accretion
flows-ADAF. We confirm the original prediction that the inner
parts of the loops are continuously advected toward the central
black hole and contribute to the growth of the large scale
magnetic field, whereas the outer parts of the loops are
continuously diffusing outward through the turbulent accretion
flow. This process of inward advection of the axial field and
outward diffusion of the return field proceeds all the way to
equipartition, thus generating astrophysically significant
magnetic fields on astrophysically relevant timescales. We confirm
that there exists a critical value of the magnetic Prandtl number
between unity and 10 in the outer disk above which the Cosmic
Battery mechanism is suppressed.
\end{abstract}


%% From the front matter, we move on to the body of the paper.
%% In the first two sections, notice the use of the natbib \citep
%% and \citet commands to identify citations.  The citations are%% tied to the reference list via symbolic KEYs.
%%The KEYcorresponds
%% to the KEY in the \bibitem in the reference list below. We have
%% chosen the first three characters of the first author's name plus
%% the last two numeral of the year of publication as our KEY for
%% each reference.

\section{Introduction}

The origin of astrophysical magnetic fields remains an open issue
of modern astrophysics. Ludwig Biermann, in his famous paper,
proposed a mechanism for the generation of large-scale electric
currents based on the thermoelectric effect (Biermann~1950). This
mechanism leads to magnetic fields that are usually quite weak
initially and require subsequent dynamo amplification to reach
astrophysically relevant magnitudes. Some years ago we proposed a
mechanism alternative to that of the Biermann Battery (Contopoulos
\& Kazanas 1998, hereafter CK, Contopoulos et al. 2006), the so
called Cosmic Battery. We posited that the aberration of the
radiation field experienced by the electrons of an accretion flow
around an astrophysical source (a generalization of the well known
Poynting-Robertson effect; Poynting~1903, Robertson~1937, Lamb \&
Miller~1995, Bini et al.~2009, Koutsantoniou \& Contopoulos~2014)
generates toroidal electric currents sufficiently large to support
poloidal magnetic fields that under certain astrophysically
plausible circumstances will grow to equipartition values. The
aberrated radiation force implies a direct coupling between
accretion flow vorticity, disk plasma diffusivity and the large
scale ordered magnetic field which may have directly observable
implications (Contopoulos et al.~2009).
%Note that the Poynting-Robertson battery effect is
%a large-scale effect that is
%independent of the presence of small-scale magnetic fields and
%magnetorotational instabilities in the accretion disk.

The magnetic field loops generated by the action of the Cosmic
Battery around the inner edge of the accretion disk (where
radiation pressure and aberration effects are maximal) are
anchored on different radii of the inner accretion disk, thus they
become twisted in the azimunthal direction by the differential
rotation of the flow. As their twisting relaxes in the vertical
direction, the loops open up and separate into an inner component
and an outer component (the return field) that threads the disk
further away from the axis. The poloidal fields of the two
components are in opposite directions, the inner one is parallel
and the outer one is antiparallel to the angular velocity vector.
On top of that, an overall helical magnetic field is produced when
the footpoints of the initially poloidal magnetic field loops are
dragged by the rotating disk plasma, with the inner toroidal
magnetic field pointing in the direction opposite to that of the
disk rotation in the northern hemisphere of the disk, and along
the direction of the disk rotation in the southern hemisphere.
According to the Cosmic Battery, reversing the observer's
hemisphere, or equivalently, the direction of disk rotation,
reverses the polarity of the axial field but leaves the direction
of the toroidal field unchanged in the observer's sky. This can be
tested in radio VLBI jets, where the direction of the toroidal
field can be directly inferred from measurements of Faraday
rotation gradients transverse to the jet axis (Blandford 1993,
Reichstein \& Gabuzda~2012, Murphy \& Gabuzda~2013). Observations
of the toroidal fields revealed from such measurements in pc-scale
jets from active galactic nuclei show that, in most cases, the
axial electric current in the VLBI core jet flows toward the
center (Contopoulos et al. 2009). The probability that this
asymmetry came about by chance was calculated to be less than
$1$\%. This observational result supports the hypothesis that the
universe is seeded by cosmic magnetic fields that are generated in
active galactic nuclei via the mechanism of the Cosmic Battery,
and are subsequently injected in intergalactic space by the jet
outflows.

In the original Cosmic Battery paper (CK) we studied the
Poynting-Robertson effect of a central isotropic radiation field
on the evolution of the magnetic field in a thin accretion flow.
We followed the evolution of the magnetic field by solving the
induction equation in 1D under the assumption that the generated
magnetic field is purely vertical. We extended this analysis in
axisymmetric configurations (2D; Contopoulos et al.~2006,
Christodoulou et al.~2008) under the assumption that the accretion
disk is connected to a steady-state force-free magnetosphere above
and below it. Our work faced some criticisms. The main argument
against the Cosmic Battery consists of the misconception that the
poloidal magnetic field loops advected to the center saturate the
magnetic field to values well below equipartition
(Bisnovatyi-Kogan et al.~2002). This criticism does not take into
account the fact that because the loops are stretched by the
differential rotation, they open up vertically, and only the inner
part of the loop is advected inward. In most cases of
astrophysical significance, the outer part of the loop diffuses
outward because astrophysical accretion disks are turbulent and
viscous, thus also diffusive, and the field growth does not
saturate.

We decided to extend our previous axisymmetric simulations to
realistic astrophysical scenaria. We consider general hot
accretion flows around a spinning black hole. The main new element
of our present simulations is that we now follow the
time-dependent evolution of the force-free magnetospheric field.
As we will see, the magnetospheric field is twisted by the
differential rotation in the disk and the central spinning black
hole. The inner part is advected toward the black hole horizon,
and the outer part diffuses through the outer accretion disk. We
obtained several animations (movies) of the field evolution which
the interested reader may find in the web (e.g.
\verb+http://youtu.be/keF0V8xwons+). Our present analysis is
Newtonian (i.e. non-general relativistic), but we expect it to be
also valid qualitatively when generalized in general relativity.

In \S~2 we present a brief overview of various disk models, and
outline the main physical processes that we consider in our
integration of the induction equation. In \S~3 we discuss how we
couple the disk with a force-free magnetosphere above and below
it. In \S~4 we discus the fundamental importance of the magnetic
Prandtl number with two characteristic examples, one that exhibits
continuous magnetic field growth, and one that exhibits saturation
of the mechanism. Finally, in \S~5 we discuss various
astrophysical examples of ADAF disks and reach our conclusions.

\section{ADAF Disks and the Induction Equation}

Black hole accretion is a physical process that provides the
primary source which powers active galactic nuclei (hereafter
AGN), black hole X-ray binaries (hereafter XRB) and possibly also
gamma-ray bursts (hereafter GRB). Black hole accretion flows are
divided into two categories, cold and hot. Cold accretion flows
are described by the standard thin disk model (Shakura \&
Sunyaev~1973, Novikov \& Thorne~1973, Lynden-Bell \&
Pringle~1974). In this case the disk is geometrically thin but
also optically thick and quite cold. The thin disk  model is
applicable when the mass accretion rate is somewhat below the
Eddington rate. In this paper we study the cosmic battery in the
general case of a hot accretion flow around a rotating black hole.

The first model for hot accretion was obtained by Shapiro et
al.~(1976). In this case the plasma is more tenuous and hotter
than the plasma in a thin disk, and is also optically thin. The
main characteristic of this model is the introduction of a
two-temperature accreting plasma where the ions have higher
temperature than the electrons. This model succeeded for the first
time in explaining the hard X-ray emission seen in black hole
sources. Unfortunately, it is thermally unstable, and therefore it
cannot be a good representation for astrophysical accretion disks.
Ichimaru~(1977) was the first who studied the important role of
advection in hot accretion flows. He pointed out that, in certain
regimes, the viscously dissipated accretion energy can go
preferentially into heating the accretion flow than into radiation
(Ichimaru 1977, Rees et al. 1982, Yuan \& Narayan 2014). This is
the central element of advection dominated accretion flows
(hereafter ADAF). The discovery of ADAF solutions (Narayan \& Yi
1994, Narayan \& Yi 1995a, Narayan \& Yi 1995b, Abramowicz et
al.~1995, Chen et al. 1995, Yuan \& Narayan 2014) gave rise to the
application of ADAF models in various black hole systems. These
systems include the supermassive black hole in the Galactic
center, Sagittarius A, low-luminosity AGNs and  black hole X-ray
binaries (e.g. Narayan et al. 1998, Abramowicz \& Fragile 2013,
Narayan \& McClintock 2008, Ho 2008, Quataert 2001, Lasota 1999).

ADAF models are characterized by the following six equations for
the radial velocity $\upsilon_{r}$, the azimuthal velocity
$\upsilon_\phi$, and the isothermal sound speed $c_s$ (Narayan \&
Yi 1994),
\begin{eqnarray}
\upsilon_{r} & = & -\frac{3a}{5+2\epsilon/f}\upsilon_{K}
\nonumber\\
\upsilon_{\phi} & = &
\frac{2\epsilon/f}{5+2\epsilon/f}\upsilon_{K}
\nonumber\\
c_{s} & = & \left(\frac{2}{5+2\epsilon/f}\right)^{1/2}
\upsilon_{K}\ , \label{ADAF}
\end{eqnarray}
where, $\upsilon_{K}\equiv (GM/r)^{1/2}$ is the Keplerian
velocity, $M$ is the mass of the central compact object/black
hole, and $c_s$ is the speed of sound.
$\epsilon=\frac{(\frac{5}{3}-\gamma)}{(\gamma-1)}$ where $\gamma$
is the  ratio of  specific  heats. $\gamma=\frac{4}{3}$ in a
relativistic plasma. The advection parameter $0<f\leq 1$ measures
the fraction of the viscously generated heat that is advected with
the flow. When $f\rightarrow 1$, most of the generated heat is
advected with the flow, whereas when $f\rightarrow 0$ the
accretion flow radiates away all of its generated heat. In that
case we end up with a standard thin Shakura-Sunyaev disk (Shakura
\& Sunyaev~1973) with accretion constant $a\approx 0.1$ (this
value corresponds to accretion disks around black holes;
Komissarov 2006). We will henceforth work in spherical coordinates
$(r,\theta,\phi)$. Notice that the expression for the sound speed
is needed in order to determine the scale height of the disk
$h\equiv r c_s/\upsilon_K$, or equivalently the polar angle of the
surface of the disk
\begin{equation}
\theta_{\rm disk}\equiv \cos^{-1}\left(\frac{h}{r}\right)=
\cos^{-1}\left(\frac{c_s}{\upsilon_K}\right)\ . \label{height}
\end{equation}

Eqs.~(\ref{ADAF}) are non-relativistic. On the other hand, the
Cosmic Battery is most effective around the inner edge of the
accretion disk around an astrophysical black hole which is
believed to coincide with the radius of the innermost stable
circular orbit (hereafter ISCO). We do not plan to perform a full
general relativistic numerical simulation in that region. Instead,
we extend our Newtonian formulation down to the black hole
horizon. We believe that, the exact details of the accretion flow
in that region are not particularly important. As we will see
below, what is most important is the fact that {\em inside the
ISCO, the flow transitions from being mostly Keplerian, thus also
turbulent, viscous, and diffusive, to being mostly freely falling,
thus also laminar, non-viscous, and non-diffusive.}

Our aim in this paper is to investigate the action of the Cosmic
Battery in realistic astrophysical accretion disks. We will thus
assume the above velocity profile inside the disk (eqs.~1) and
integrate the induction equation (eq.~\ref{Induct} below) in order
to follow the evolution of the disk magnetic field in the presence
of the Cosmic Battery radiation source. In other words, we will
assume that the field is passive and does not affect the velocity
field in the disk.

The induction equation in the disk ($\theta_{\rm disk}\leq
\theta\leq 90^\circ$) can be written as
\begin{equation}
\frac{\partial {\bf B}_{\rm disk}}{\partial t}= -c\nabla\times
\left(E_{\rm CB}\hat{\phi}-\frac{\upsilon}{c} \times {\bf B}_{\rm
disk}+ \eta\nabla\times {\bf B}_{\rm disk}\right) \label{Induct}
\end{equation}
where ${\bf B}_{\rm disk}$ is the magnetic field, and $\eta$ is
the magnetic diffusivity in the disk. The Cosmic Battery
electromotive source term $E_{\rm CB}$ is equal to the aberrated
radiation force on the electrons in the azimuthal direction
$F_{\rm rad}|_{\phi}$ divided by the proton charge $e$. In our
original paper, we considered a simplified central radiation field
with luminosity $L$, in which
\begin{equation}
E_{\rm CB}=\frac{F_{\rm rad}|_{\phi}}{e}=-\frac{L\sigma_T}{4\pi c
er^2}\frac{\upsilon_\phi}{c} \label{simplefrad}
\end{equation}
where, $\sigma_T$ is the electron Thompson cross section. The
correct expression for the aberrated radiation force requires a
complex general relativistic calculation of the radiation field as
felt by the moving electron, taking into consideration the plasma
optical depth and the fact that the source of radiation is the
accretion disk itself. This calculation has been performed for the
first time by Koutsantoniou \& Contopoulos~(2014) for an optically
thick flow with a thermally emitting surface. The case of an
optically thin disk is much more complex and is currently under
investigation. In order to proceed, we will assume a slightly more
general phenomenological expression
\begin{equation}
E_{\rm CB}=-\frac{L\sigma_T}{4\pi
ecr^2}\frac{\upsilon_\phi}{c}\times \left\{ \begin{array}{ll} 1\
\mbox{or}\ e^{-\frac{(r-r_{\rm ISCO})^2}{2 r_{\rm ISCO}^2}}
& \mbox{if}\ r\geq r_{\rm ISCO}\\ \\
e^{-\frac{(r-r_{\rm ISCO})^2}{2 (0.2 r_{\rm ISCO})^2}} &
\mbox{if}\ r< r_{\rm ISCO}
\end{array}
\right.
\label{CB}
\end{equation}
which incorporates the main features of a realistic radiation
force: the buildup of the radiation field as we approach the ISCO,
%the optical depth of the flow beyond a certain distance,
the Poynting-Robertson (radiation aberration) effect, and the
$1/r^2$ drop-off with distance of the radiation pressure in an
optically thin medium. We also have the option to account for some
long optical depth (on the order of $r_{\rm ISCO}$) in the disk
(second expression in top line of eq.~\ref{CB} above). As we said
above, a more detailed general relativistic calculation of the
radiation force that takes into account the physical optical depth
of the accretion flow is under way (Koutsantoniou \& Contopoulos
in preparation).

The magnetic diffusivity $\eta$ can be expressed as
\begin{equation}
\eta=\left\{ \begin{array}{ll}
\frac{arc_{s}^{2}}{\upsilon_{K}{\cal P}_m}
& \mbox{if}\ r\geq r_{\rm ISCO}\\ \\
0 & \mbox{if}\ r< r_{\rm ISCO}
\end{array}
\right.
\end{equation}
(see for example Lovelace et al. 2009). The magnetic diffusivity
is therefore a function of the magnetic Prandtl number ${\cal
P}_m$ which itself is in general a function of distance in the
accretion flow. In the present work we will assume that ${\cal
P}_m$ is roughly a constant within the spatial extent of our
numerical simulations, but very quickly becomes infinite (i.e.
$\eta$ drops very quickly to zero) inside $r_{\rm ISCO}$ as the
flow very quickly changes character from quasi-Keplerian to almost
free fall on the black hole horizon. This transition is related to
the fact that as the flow rotation decreases, the shear decreases,
and the flow becomes laminar inside the ISCO because matter is
almost in free fall there. The laminar flow implies that the
turbulence responsible for both the flow viscosity and the
magnetic diffusivity ceases.

\subsection{Numerical setup}

We solve our equations in a numerical grid with uniform
latitudinal spacing $\delta\theta$, linearly increasing radial
spacing $\delta r=3r\delta\theta$, and a finite radial extent
$r_{\rm in}\leq r\leq r_{\rm out}$, where $r_{\rm in}=r_{\rm
ISCO}/3$, and $\theta_{\rm min}\leq \theta\leq 90^\circ$. Although
our formalism is Newtonian, the inner radial boundary $r_{\rm in}$
of our grid is taken to represent the black hole horizon. An inner
polar cutoff of a few degrees $\theta_{\rm min}\sim 5^\circ$ is
implemented in order to avoid coordinate system singularities
around the axis. Our standard $r\times \theta$ grid resolution is
$87 \times 90$ respectively, and our numerical grid extends
radially out to $r_{\rm out}=25 r_{\rm ISCO}$. We have also run
selected simulations with double grid resolution, namely $174
\times 180$, and found agreement within less than 5\%. We have
thus concluded that our present numerical results have converged.
We have also run simulations on a uniform $(\delta r,
\delta\theta)$ grid and obtained essentially the same results.

The magnetic field is generated by the Cosmic Battery and is not
brought in from large distances, therefore, the time integration
of eq.~(\ref{Induct}) begins with
\begin{equation}
{\bf B}_{\rm disk}(t=0)=0\ .
\end{equation}
Also, we set $B_r=B_\phi=0$ along the equator $\theta=90^\circ$,
and implement free inner and outer boundaries at $r_{\rm in}$ and
$r_{\rm out}$ respectively. What is most important is that the
magnetic field along the surface of the disk is matched to the
magnetospheric field, namely
\begin{equation}
{\bf B}_{\rm disk}(\theta_{\rm disk})={\bf B}_{\rm MS}(\theta_{\rm
disk})\ . \label{Bsurface}
\end{equation}
The magnetospheric field ${\bf B}_{\rm MS}(r,\theta)$ for
$\theta\leq \theta_{\rm disk}$ is obtained in an independent
calculation that we are now going to discuss.

\begin{figure*}[t]
\centering
\includegraphics[trim=0cm 0cm 0cm 0cm,
clip=true, width=16cm, angle=0]{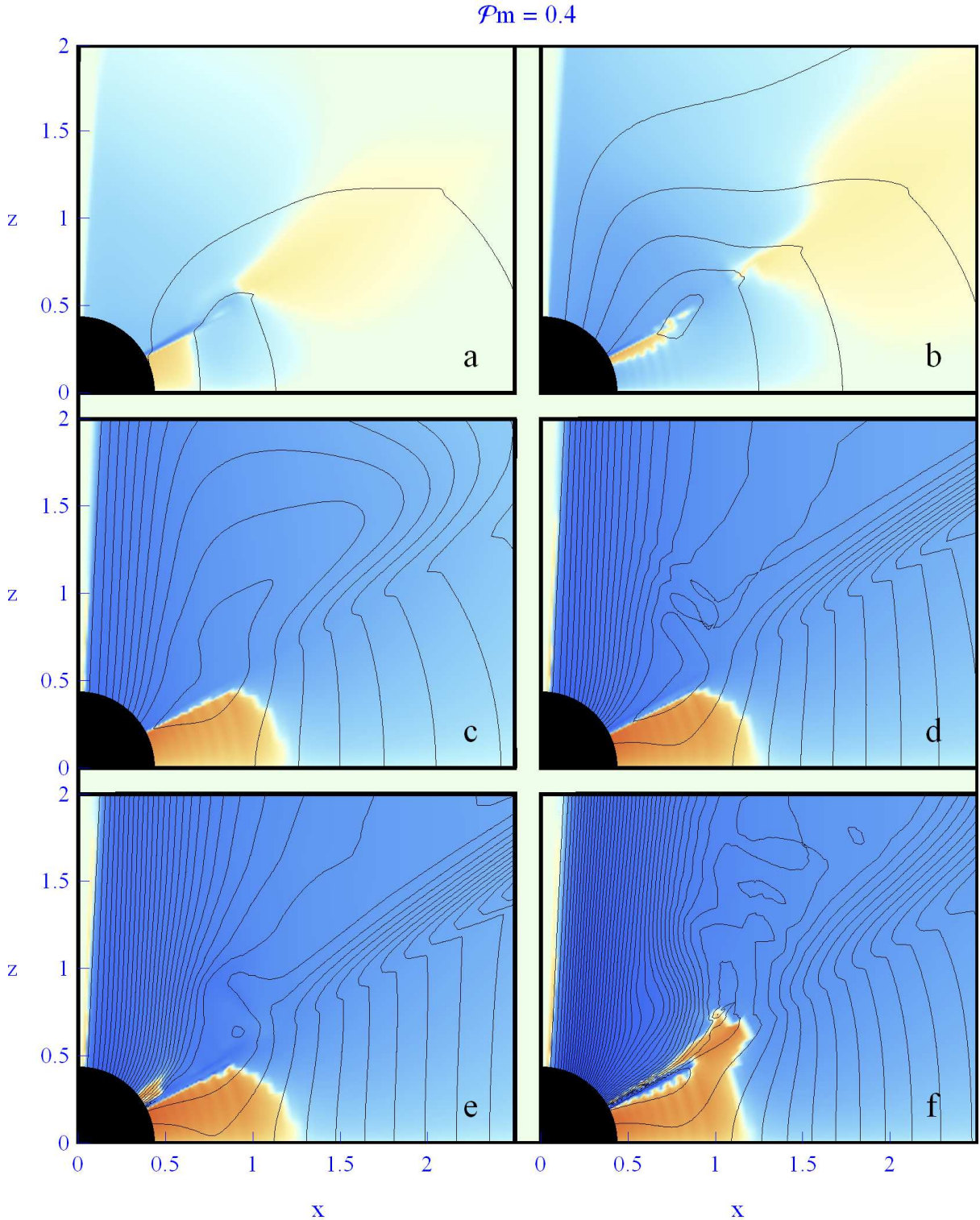} \caption{Poloidal
magnetic field lines at times $t_{\rm acc}$ (a), $3t_{\rm acc}$
(b), $8.6t_{\rm acc}$ (c), $15.7t_{\rm acc}$ (d), $23.7t_{\rm
acc}$ (e), and  $38.4t_{\rm acc}$ (f), in the standard case of a
diffusive accretion disk with Prandtl number ${\cal P}_m=0.4$.
Length scales in unit of $r_{\rm ISCO}$. Colors represent the
value and sign of $B_\phi$ in arbitrary units (blue: negative,
orange: positive). The surface of the disk lies at
$\theta=60^\circ$. Poloidal field loops are continuously generated
around $r_{\rm ISCO}$. The inner footpoints of the field lines are
advected together with infalling matter to the central object, and
magnetic flux of one polarity is accumulated on the central object
at a steady rate. The disk is threaded by the return field
polarity. The outer footpoints are diffusing outwards inside the
accretion disk, and because of the differential rotation of their
footpoints, the loops open up. A steady configuration is
established where poloidal loops are continuously generated around
the inner edge of the disk and open up to infinity.} \label{stand}
\end{figure*}

\section{The Force-Free Magnetosphere}

Whenever the magnetic field generated by the Cosmic Battery in the
disk reaches the surface it will extend into the disk atmosphere,
the so-called magnetosphere. In order to follow the evolution of
the magnetic field in the accretion flow, we must also numerically
evolve the field in the magnetosphere just above the surface of
the disk. It is natural to assume that, near the surface of the
disk, the magnetosphere attains a force-free configuration, in
analogy to the solar corona and the pulsar magnetosphere. The
equations that describe the field evolution are those of
force-free electrodynamics (FFE; Gruzinov~1999, Blandford~2002),
namely Maxwell's equations in the presence of electric charges and
currents,
\begin{eqnarray}
\frac{\partial {\bf B}_{\rm MS}}{\partial t} & = & -c\nabla\times
{\bf E}_{\rm MS}
\nonumber\\
\frac{\partial {\bf E}_{\rm MS}}{\partial t} & = & c\nabla\times
{\bf B}_{\rm MS}
-4\pi {\bf J}\nonumber\\
\nabla\cdot {\bf B}_{\rm MS} & = & 0\nonumber\\
\nabla\cdot {\bf E}_{\rm MS} & = & 4\pi \rho_e\ , \label{FFE1}
\end{eqnarray}
supplemented by the force-free and ideal MHD conditions
\begin{eqnarray}
\rho_e {\bf E}_{\rm MS} + {\bf J}
\times {\bf B}_{\rm MS} & = & 0\\
{\bf E}_{\rm MS}\cdot {\bf B}_{\rm MS} & = & 0\ . \label{FFE2}
\end{eqnarray}
Here, $\rho_e$, ${\bf J}$ are the magnetospheric electric charge
and current densities respectively. The above formulation is
obviously non-general relativistic, yet we believe that the
simulations performed in this study can give an overall impression
about how the Cosmic Battery mechanism works. Notice that recent
general relativistic simulations of force-free disk magnetospheres
(Parfrey et al.~2015) show a magnetic field evolution similar to
the one we obtain here (magnetic field loops becoming twisted and
opening up due to the differential rotation of the flow in the
disk, current sheet and plasmoid formation, etc.). We plan to
extend our numerical calculations in general relativity in a
forthcoming paper.

\subsection{Numerical setup}

The numerical code we employ is the 3D code first developed in
Kalapotharakos \& Contopoulos~(2009), re-written in spherical
coordinates $(r,\theta,\phi)$ by Contopoulos~(2013). We have
lowered the dimensions to 2D by assuming axisymmetry. As in the
disk interior, we implement an inner free boundary at $r=r_{\rm
in}$ in this region too. In practice, we allow for any magnetic
flux that is generated inside the disk and reaches the inner
magnetospheric boundary (which represents the black hole horizon),
to be freely distributed along that boundary. Beyond a certain
distance $r_{\rm PML}< r_{\rm out}$, we assume an absorbing
non-reflecting outer boundary in the form of a special
implementation of an axisymmetric Perfectly Matched Layer (PML;
discussed in detail in Kalapotharakos \& Contopoulos~2009), namely
\begin{eqnarray}
\frac{\partial {\bf B}_{\rm MS}}{\partial t} & = & -c\nabla\times
{\bf E}_{\rm MS}
-\sigma(r) {\bf B}_{\rm MS}\nonumber\\
\frac{\partial {\bf E}_{\rm MS}}{\partial t} & = & c\nabla\times
{\bf B}_{\rm MS} -\sigma(r) {\bf E}_{\rm MS}\ , \label{PML}
\end{eqnarray}
where
\begin{equation}
\sigma(r)=\sigma_o \frac{c}{r_{\rm ISCO}}\left(\frac{r-r_{\rm
PML}}{r_{\rm out}-r_{\rm PML}}\right)^3\ . \label{sigma}
\end{equation}
We have taken $\sigma_o=100$, and $r_{\rm PML}=17r_{\rm ISCO}$.
These parameters, as well as the exponent in eq.~(\ref{sigma})
have been chosen empirically. Notice the similarity with the
equivalent expression in Cerutti et al.~(2014). We have run
several simulations at various radial resolutions and various
values of the outer boundary $r_{\rm out}$ and the results are
similar (the simulations have converged). For most of the results
we obtain in the present paper, the outer boundary was placed at
$25 r_{\rm ISCO}$.

As we said in the previous section, the evolution of the magnetic
fields in the two domains (the disk and the magnetosphere) are
related. It is important to notice here that there is an important
complication that enters in the matching between the two domains,
and this has to do with the abrupt transition between the dense
turbulent accreting flow and the almost empty ionized (probably
also evaporating and outflowing) magnetosphere.  This is very
similar to the well known transition region in the sun where,
within a very thin layer of about one hundred kilometers, the
temperature rises by almost two orders of magnitude to about one
million degrees, and the matter density drops by about one order
of magnitude. The solar transition region is the subject of
ongoing intense theoretical and numerical investigations. The
astrophysical disk transition region is at least equally
complicated since it is not only the base of the disk corona, but
also the origin of purported disk winds and outflows (e.g.
Li~1995, Ferreira \& Pelletier~1995).  is not the aim of the
present study to solve in detail for the structure of the flow and
magnetic field in this region. In practice, we define a transition
region of two $\theta$-grid zones above $\theta_{\rm disk}$. In
that region we set
\begin{equation}
{\bf B}_{\rm MS}={\bf B}_{\rm disk}\ . \label{transition1}
\end{equation}
This matching allows for magnetic field loops that reach the disk
surface zones to escape to the magnetosphere. We also set the
field lines in that region in rotation by introducing a
magnetospheric poloidal electric field
\begin{equation}
{\bf E}_{\rm MS}(\theta_{\rm
disk})=-\frac{\upsilon_\phi}{c}\hat{\phi}\times {\bf B}_{\rm
disk}(\theta_{\rm disk})\ .
\end{equation}
Finally, we slightly modify the induction equation that we solve
for the disk material in that region by introducing an extra
phenomenological term that mimics various complex physical effects
taking place in the surface layers of astrophysical accretion
disks such as surface convection, buoyancy of magnetic field
loops, and the fact that the surface layers are highly ionized
(thus also fully conducting) due to cosmic ray irradiation, namely
\begin{equation}
\frac{\partial {\bf B}_{\rm trans}}{\partial t}= -c\nabla\times
\left(E_{\rm CB}\hat{\phi}+\frac{1}{200}\hat{\theta} \times {\bf
B}_{\rm trans}\right)\ , \label{transition}
\end{equation}
where, ${\bf B}_{\rm trans}$ is the disk magnetic field in the.
The introduction of this transition layer effectively shields the
disk interior: a) it prevents any generated magnetospheric field
from `entering' the disk from above (especially in the case of
turbulent high $\eta$ flow conditions), and b) it `pushes outward'
(in the $\theta$ direction) at a very small fraction of the speed
of light any magnetic field loop that enters it from below. We
implemented different fractions ($1/200$, $1/20$, etc.) and the
results were qualitatively similar. transition region. As we said
above, we are not claiming that we study in detail the
disk-magnetosphere transition region. This is the reason we opted
for a minimal two $\theta$-zone layer that only serves the purpose
of shielding the disk interior from the disk magnetosphere. Notice
that Parfrey et al.~(2015) ignored the important physical
significance of a transition layer by directly coupling their
magnetosphere to a prescribed distribution of the magnetic field
on the surface of the accretion disk, {\em without} solving for
the magnetic field distribution in the disk interior.

The inner boundary (which represents the black hole horizon) also
introduces a rigid body rotation expressed through
\begin{equation}
{\bf E}_{\rm MS}(r_{\rm in},\theta)=-\frac{\upsilon_\phi(r_{\rm
in},\theta_{\rm disk})\sin(\theta)}{c\sin(\theta_{\rm
disk})}\hat{\phi}\times {\bf B}_{\rm MS}(r_{\rm in},\theta)
\label{BHrotation}
\end{equation}
for $\theta \leq \theta_{\rm disk}$. The latter turns out not to
be important in the determination of the overall magnetospheric
field topology, and plays a role only if we are interested in the
extraction of energy from the black hole rotation (Blandford \&
Znajek~1977). As in the disk interior, the time integration of
eqs.~(\ref{FFE1}-\ref{PML}) begins with
\begin{equation}
{\bf B}_{\rm MS}(t=0)=0
\end{equation}
everywhere in the magnetosphere.

\section{The Cosmic Battery in Action}

The goal of this paper is to convince the reader that the Cosmic
Battery plays a fundamental role in the generation of the large
scale magnetic fields that thread astrophysical accretion disks
and are believed to be responsible for such diverse astrophysical
phenomena as jets, AGNs, XRBs, and possibly also GRBs. We
performed numerical integrations that couple the FFE equations in
the magnetosphere above and below the disk
(eqs.~\ref{FFE1}-\ref{FFE2}) with the induction equation that
includes the Cosmic Battery electromotive force and the flow
magnetic diffusivity in the disk (eq.~\ref{Induct}). As we said,
the coupling is performed in a thin surface transition region
through eqs.~(\ref{transition1}-\ref{transition}). Our goal is to
find out how much magnetic flux can be accumulated around the
central compact object. In particular, we are going to show under
what conditions the accumulated magnetic flux grows steadily to
equipartition, and under what conditions it saturates to a very
low value. We want to make clear that these are not MHD flow
simulations since the accretion flow is independently specified by
the ADAF equations (eqs.~\ref{ADAF}).

\begin{figure*}[t]
\centering
\includegraphics[trim=0cm 0cm 0cm 0cm,
clip=true, width=16cm, angle=0]{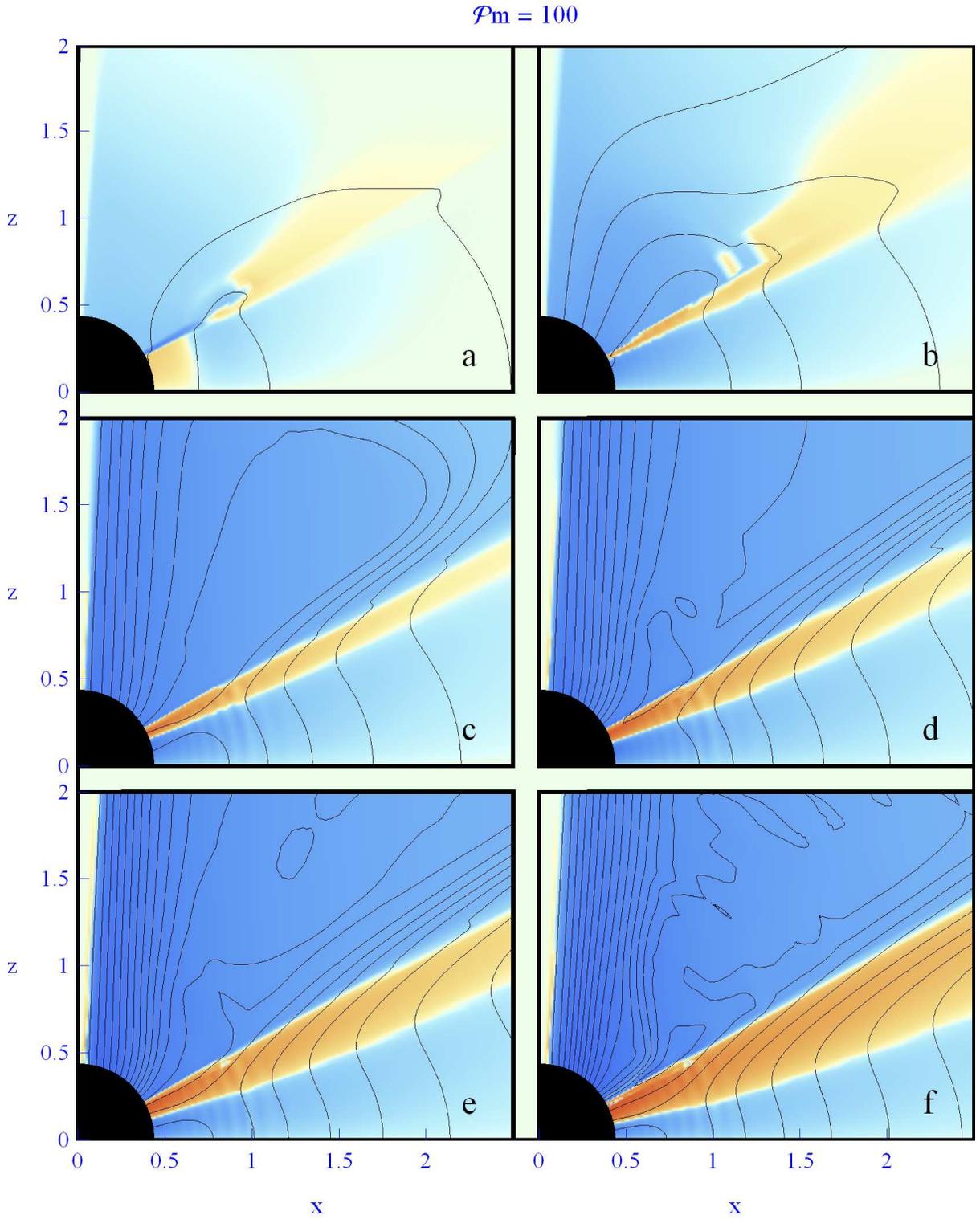} \caption{Same as
Fig.~1 for the very low $\eta$ case of a non-diffusive optically
thin accretion disk with magnetic Prandtl number ${\cal P}_m=100$.
The inner footpoints of the field lines are advected together with
infalling matter onto the central object. In the outer disk region
there is a balance between inward advection (that wants to bring
the outer footpoints to the central object), outward diffusion,
and continuous field generation due to the action of the Cosmic
Battery. The rate of growth of the accumulated magnetic field
keeps decreasing and eventually saturate.} \label{low}
\end{figure*}

\begin{figure}[t]
\centering
\includegraphics[trim=0cm 0cm 0cm 0cm,
clip=true, width=8cm, angle=0]{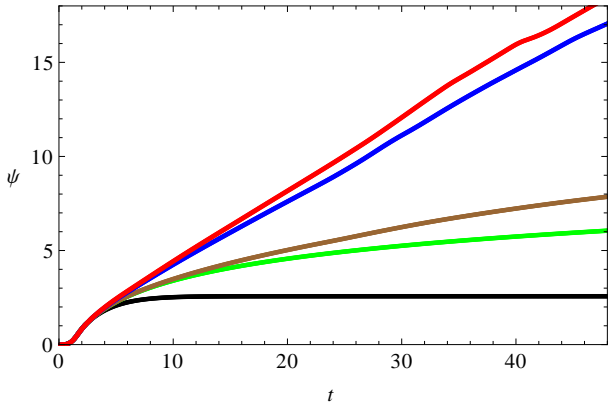} \caption{Evolution of
the magnetic flux $\Psi$ accumulated on the central object (in
units of $\Psi_{\rm acc}$) for various values of the magnetic
Prandtl number (from top to bottom: ${\cal P}_m=0.2, 0.4, 1, 10,
100$, and $100$ with finite optical depth in the disk). Time in
units of $t_{\rm acc}$} \label{fig1a}
\end{figure}

\begin{figure*}[t]
\centering
\includegraphics[trim=0cm 0cm 0cm 0cm,
clip=true, width=16cm, angle=0]{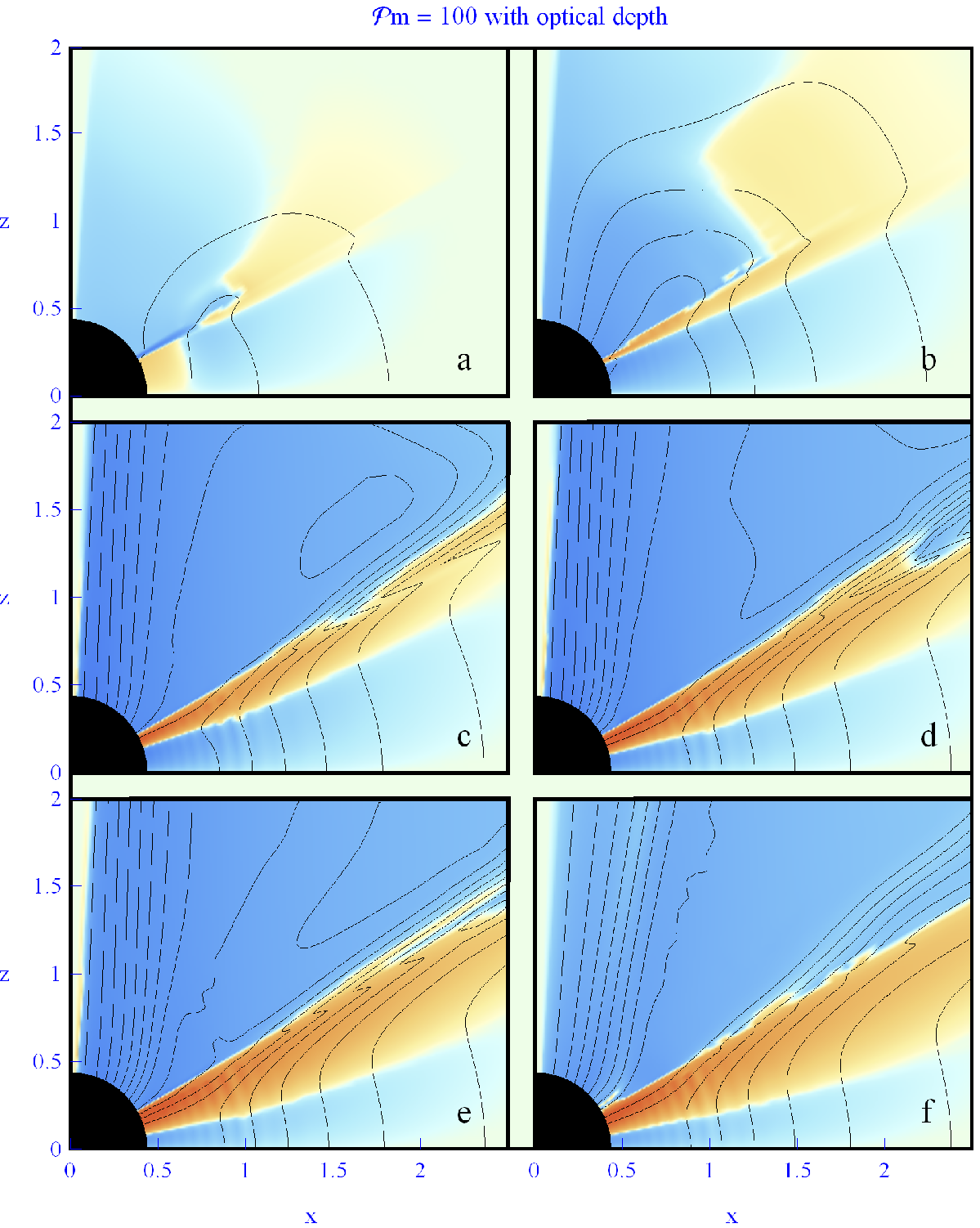} \caption{Same as
Fig.~2 but now an optical depth of order $r_{\rm ISCO}$ is
included in the disk and the Cosmic Battery dies out beyond a few
times $r_{\rm ISCO}$. Saturation is reached very quickly and
beyond time $t\approx 5 t_{\rm acc}$ no more magnetic flux is
brought to the central object.} \label{lowdepth}
\end{figure*}

In this section, we are going to highlight the fundamental role of
the accretion disk magnetic diffusivity in activating (or under
different conditions saturating) the Cosmic Battery. This point,
already discussed in the original CK presentation, has been missed
by Bisnovatyi-Kogan et al.~(2002) in their criticism of the Cosmic
Battery. We will thus now focus on two particular simulations, one
with moderate and one with low magnetic diffusivity. The first
simulation corresponds to a magnetic Prandtl number ${\cal
P}_m=0.4$. We will call this case the {\em `standard'} case. In
the second simulation we set ${\cal P}_m=100$. We will call this
case the {\em `very low} $\eta$' case. Notice that, in order to
minimize the number of free parameters in the problem, we fix the
accretion constant $a=0.1$, and the advection parameter $f=1$ (the
latter value corresponds to a strongly advection-dominated flow).
We will now compare the evolution of our two characteristic
simulations.

In the beginning, the Cosmic Battery term (eq.~\ref{CB}) generates
poloidal magnetic loops around $r_{\rm ISCO}$ in both cases. The
inner footpoints of the field lines are advected by disk matter
that falls into the central object. Furthermore, as each loop
enters the magnetosphere it starts to inflate due to the
differential rotation between its inner and outer footpoints on
the surface of the disk. This can be seen in Figs.~1a and 2a which
are very similar. As the mechanism continues to operate, more and
more field lines are brought to the central object. The
similarities end beyond about one accretion time defined as the
time needed to traverse a distance equal to $r_{\rm ISCO}$ at the
ISCO accretion speed, namely
\begin{eqnarray}
t_{\rm acc} & \equiv & \frac{r_{\rm ISCO}}{\upsilon_r(r_{\rm
ISCO})}\nonumber \\
 & = &
 \left(\frac{5+2\epsilon/f}{3a}\right)
 \frac{r_{\rm ISCO}}{v_K(r_{\rm ISCO})}\nonumber\\
 & \approx &
 1.7\times 10^{-2}
 \left(\frac{M}{10 M_{\odot}}\right)\ \mbox{sec}
 \ . \label{tacc}
\end{eqnarray}
What differentiates the two cases is the behavior of the
outer footpoints beyond that time.

\subsection{The standard case}

In the standard case, the outer footpoint of a poloidal field line
generated by the Cosmic Battery begins to diffuse outward inside
the accretion disk. Because of the difference in the rotation
between the two footpoints, the field lines wind up around the
central object. They inflate in the magnetosphere above the
accretion disk and sustain a connection between the central object
and the disk (Fig.~1b). New loops are continuously generated
around the position of the ISCO (Fig.~1b-f). As the outer
footpoints are continuously moving outward, the loops expand more
and more. Their inner and outer parts never disconnect, and the
sign of the magnetospheric $B_{\phi}$ component remains unchanged
(negative). Notice that although $B_{\phi}$ develops highly
variable structure on small scales (as can be seen in Figs.~1e \&
f), it does not affect the large scale field structure, and the
numerical integration is stable. The return polarity diffuses
outward and the field in the disk remains at very low
(sub-equipartition) values. Such low magnetic field cannot be
responsible for a disk MHD wind (e.g. Li~1995, Ferreira \&
Pelletier~1995). This is an interesting point that may be
investigated more in the future.

If the central object is rotating, then the field lines of one
polarity that have already reached it can extract its rotational
energy (Blandford \& Znajek~1977, Nathanail \& Contopoulos~2014).
In this paper, we are interested in showing the growth of the
magnetic field, and therefore, we are not going to discuss this
effect any further. It is interesting to notice that, for the
particular ADAF parameters of our present simulation, the inner
magnetospheric boundary is set in uniform rotation with
$\Omega(r_{\rm in})=\Omega_{\rm disk}(r_{\rm in},\theta_{\rm
disk})=0.6 c/r_{\rm ISCO}$. This corresponds to a magnetospheric
light cylinder at cylindrical radius
\begin{equation}
R_{\rm LC}\equiv \frac{c}{\Omega(r_{\rm in})}=1.67 r_{\rm ISCO}\ .
\label{LC}
\end{equation}

As shown in Fig.~3, the magnetic flux brought to the central
object grows at a steady rate which has no reason to cease before
the field reaches equipartition with the accretion flow (defined
as the limit where the accumulated magnetic field pressure
$B^2/4\pi$ becomes comparable to the ram pressure $\rho \upsilon_r
\upsilon_\phi$ of the flow). In Fig.~3, we have normalized $\Psi$
to a characteristic value $\Psi_{\rm acc}$ that corresponds to the
amount of flux accumulated interior to $r_{\rm ISCO}$ generated by
the Cosmic Battery electromotive force term in one accretion time,
namely
\begin{eqnarray}
\Psi_{\rm acc} & \equiv & \frac{\pi r_{\rm ISCO}^2 E_{\rm
CB}(r_{\rm ISCO})c}{\upsilon_r(r_{\rm ISCO})}\nonumber\\
 & = &
\left(\frac{\upsilon_\phi}{\upsilon_r}\right)\frac{L\sigma_T}{4ce}
\approx \frac{\epsilon}{6fa}\frac{L\sigma_T}{ce}\nonumber\\
& \approx & 10^{14}\ \mbox{G}\ \mbox{cm}^2 \left(\frac{L}{L_{\rm
Edd}}\right)
\left(\frac{M}{10 M_{\odot}}\right)\nonumber\\
& \approx & 0.4\ {\rm G}\ \pi r_{\rm ISCO}^2 \left(\frac{L}{L_{\rm
Edd}}\right)\left(\frac{M}{10 M_{\odot}}\right)^{-1}\ .
\label{Psiacc}
\end{eqnarray}
What is important to notice is that the magnetic field that
threads the disk interior remains always at a very low value
since, inside the ISCO it is continuously advected toward the
central boundary (the black hole horizon), whereas outside it is
continuously diffusing outward through the accretion disk. As long
as the field remains at such small values, the MRI, the physical
mechanism believed to be responsible for the disk turbulence
(Balbus \& Hawley~1992), does not saturate.

\subsection{The very low $\eta$ case}

The field evolution in the very low $\eta$ case is very different.
Beyond the accretion time $t_{\rm acc}$ needed for matter to reach
the central object, the mechanism begins to saturate. As we
pointed out above (and as can be seen in Fig.~2a), at times up to
a few times $t_{\rm acc}$ the evolution is the same as in the
standard case, namely the inner field footpoints are carried by
matter falling to the central object and the outer footpoints are
rotating with the disk. However, the accretion disk is no longer
diffusive and the outer footpoints do not diffuse easily outward
inside the disk. In fact, the accretion flow wants to bring also
the outer footpoints to the central object. The whole loop would
be swallowed by the central object, but this is not the case since
new magnetic field loops are continuously generated by the Cosmic
Battery in the optically thin part of the disk. Due to this, the
mechanism instead saturates. We see the deviation from the
standard case already at times $t\sim 10 t_{\rm acc}$ when the
inner and outer parts of the poloidal magnetic field loops begin
to disconnect (Fig.~2b,c). Beyond that time, the field lines of
the return polarity which are anchored at larger distances on the
disk are now wound by the disk rotation at their footpoints, and
not by the differential winding between the inner and outer
footpoints of the loop (which, as we said, are now disconnected).
When this happens, the $B_{\phi}$ component of the return (outer)
field line changes sign (becomes positive), as can be seen by the
change of color near the surface of the disk. A dynamic
magnetospheric current sheet forms between the two regions of
opposite poloidal field polarity which in our very-low $\eta$ case
lies a few degrees above the surface of the disk (Figs.~2d-f). The
ongoing (numerical) reconnection forms plasmoids that outflow fast
along the current sheet. Plasmoid formation and reconnection in
the current sheet may be responsible for particle acceleration and
dissipation of energy. Magnetic flux is continuously advected on
the central object, but at a continuously reduced rate (see
Fig.~3).

We have also run simulations in which we implement a long but
finite optical depth in the disk according to the second term in
the first line of eq.~\ref{CB}, for the same value of the disk
diffusivity (low $\eta$). This effectively kills the Cosmic
Battery beyond a radial distance of a few $r_{\rm ISCO}$. In that
case, the field saturation is much clearer as can be seen in
Figs.~3 \& 4.
%This current
%sheet is a crucial integral part of the solution. In a
%steady-state, all field lines that thread the innermost boundary
%(the black hole) extend from the axis up to the magnetospheric
%current sheet.

Notice that, in the very low $\eta$ case, the saturation value of
the magnetic flux that was brought to the central object is very
low, thus, the observed current sheet activity observed in this
case is probably astrophysically insignificant. The big difference
from the standard case is that in the latter, the magnetic flux
continues to grow at a steady rate, presumably up to
equipartition.

\section{Discussion and Conclusions}

\begin{figure*}[t]
\centering
\includegraphics[trim=0cm 0cm 0cm 0cm,
clip=true, width=11cm, angle=0]{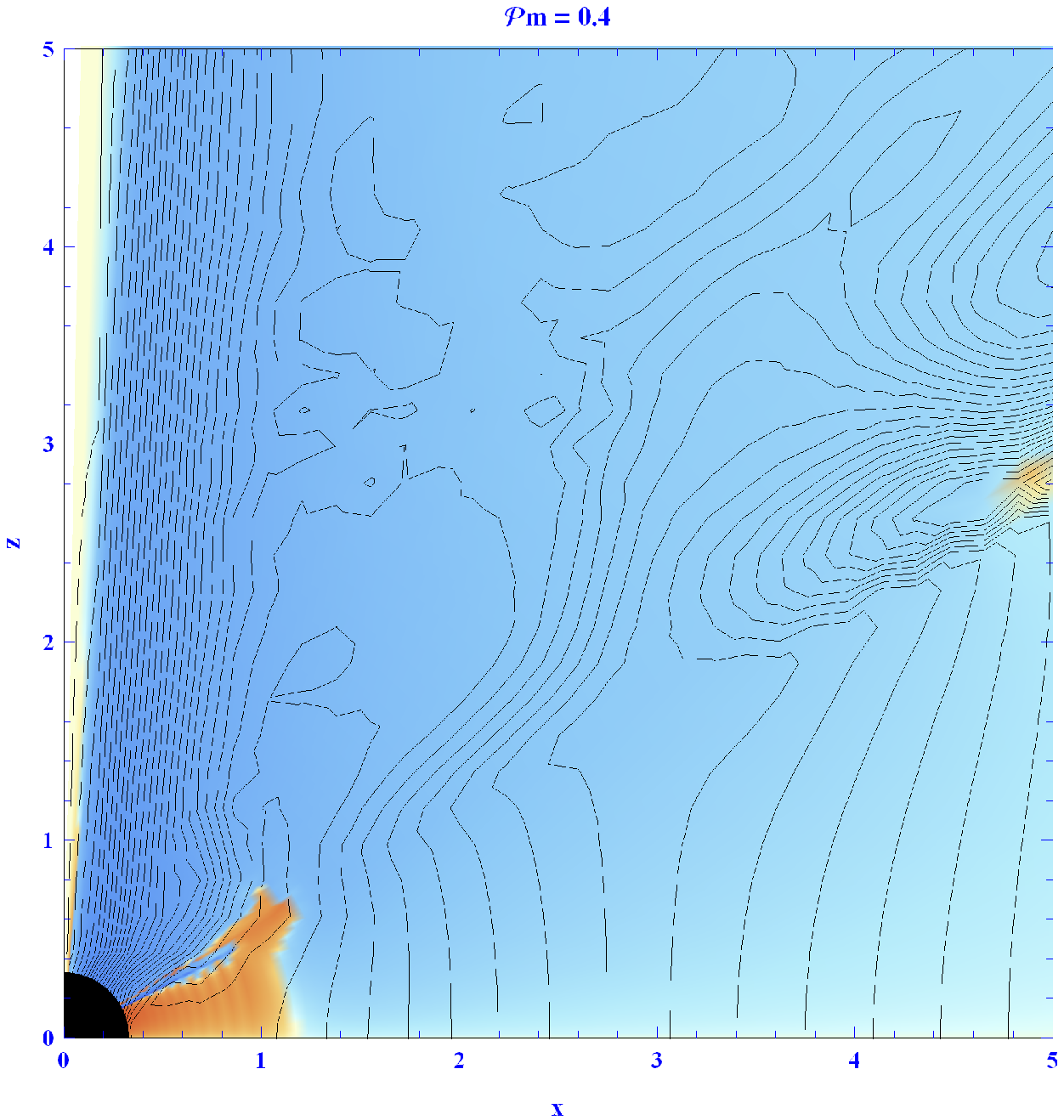}
\caption{Configuration (f) of Fig.~1 (${\cal P}_m=0.4, t=38.4
t_{\rm acc}$) shown over a larger spatial scale. The central flux
accumulation proceeds unimpeded. However, at larger distances, the
disk develops complex dynamic magnetic activity.}
\label{standlarge}
\end{figure*}

In the previous section we argued that the main factor that allows
for the steady growth of the accumulated magnetic flux is the
value of the magnetic Prandtl number ${\cal P}_{m}$ in the disk.
We run several simulations for different values of the magnetic
Prandtl number in the accretion flow (Fig.~3). In these
simulations, other physical parameters are kept constant. It is
interesting to see that in all our runs the magnetic flux advected
to the central object within a few accretion times $t_{\rm acc}$
remains almost the same (on the order of $\Psi_{\rm acc}$). Within
that short amount of time, the mechanism does not feel the
diffusivity properties of the disk. As time goes by, only
diffusive disks allow the magnetic flux to grow. As can be seen in
Fig.~3, the rate of growth is closely coupled with the magnetic
diffusivity of the disk. When the Prandtl number is low (high
diffusivity), the rate of growth is bigger. For the lowest values
that we have tried (${\cal P}_m=0.2, 0.4, 1$), the growth of the
magnetic flux is linear with similar slope after about $t_{\rm
acc}$ from the beginning of the run. As the Prandtl number
increases beyond unity (which corresponds to lower and lower
diffusivity), the rate of growth of the magnetic flux continuously
decreases with time, approaching zero asymptotically (no growth,
i.e. saturation). There seems to be a critical value of the
magnetic Prandtl number for this transition in the range $1 <
{\cal P}_{m\ {\rm crit}} < 10$. For Prandtl numbers higher than
$100$ the situation is identical to the very low $\eta$ case.

%\begin{figure}[t]
%\centering
%\includegraphics[trim=0cm 0cm 0cm 0cm,
%clip=true, width=8cm, angle=0]{fall1.eps} \caption{Evolution of
%the normalized magnetic flux $\Psi/\Psi_{\rm acc}$ accumulated on
%the central object for various values of the advection parameter
%$f$, for a fixed value of ${\cal P}_m=0.4$.} \label{fig1b}
%\end{figure}

%Another numerical experiment that we performed was to assume a
%diffusive disks with fixed Prandtl number ${\cal P}_m=0.4$ for
%which the accumulated magnetic flux grows steadily, and then try
%different values of the advection parameter $f$ (different types
%ADAFs). Notice that different advection parameters correspond to
%different speeds of sound, thus also to different disk scale
%heights $h\approx r c_s/\upsilon_K$. As we discussed before, all
%cases show the same behavior within about one accretion time.
%Beyond that time, they proceed with different growth rates. For
%$f=1$, which corresponds to strongly advection-dominated flows the
%growth rate is the lowest. As the advection parameter decreases
%down to $f=0.2$, corresponding to weak advection ADAF, the growth
%rate increases.

Our present realistic simulations of the Cosmic Battery at the
inner edge of the accretion disk around a black hole confirm that
this is an important astrophysical mechanism that may indeed
account for the generation of the large scale magnetic field that
threads astrophysical jets and disks. Under standard accretion
disk conditions (magnetic Prandtl numbers of order unity), the
Cosmic Battery generates a large scale magnetic field on the order
of $B_{\rm acc}\sim 1$~G inside the ISCO of a ten solar mass black
hole within about one accretion time on the order of a few
milliseconds. Assuming that the black hole is accreting at about
$1\%$ of its Eddington rate, in order for the accumulated magnetic
field to reach an equipartition value on the order of $10^7$~G,
one has to wait for about $10^9$ accretion times, or equivalently
for about $t_{\rm eq}\sim 10^7$~seconds (a few months). The above
numbers obviously scale with mass as can be seen in Table~1.

We have shown that the Cosmic Battery does not increase the value
of the magnetic field that threads the accretion disk. The
magnetic field remains several orders of magnitude below
equipartition, and does not inhibit the action of the MRI in the
disk. Of course, as the magnitude of the accumulated field inside
the ISCO approaches equipartition, the mechanism will saturate,
and the accumulated field will probably escape outward through the
disk due to magnetic Rayleigh-Taylor instabilities around its
inner edge (Tchekhovskoy et al. 2010, Contopoulos \& Papadopoulos
in preparation). This is interesting because when that happens,
one expects to also see magnetically driven outflows (winds) from
the disk. In other words, unless the accumulated field approaches
equipartition around the inner edge of the disk (and thus becomes
unstable to magnetic Rayleigh-Taylor instabilities), we do not
expect to see any strong wind emanating from the disk. This is
contrary to what is expected in the scenario where the magnetic
field is brought in from large distances through the disk, and
thus threads the disk and generates strong steady disk winds. It
is interesting that, although the overall growth of the generated
and accumulated magnetic flux proceeds as described above, the
return field that threads the accretion disk at large distance
beyond $r_{\rm ISCO}$ shows a very complex behavior that results
from the interaction between the accretion flow, the force-free
magnetosphere and the surface transition region (Fig.~5). Such
dynamic behavior is a far cry from the smooth steady-state
configurations obtained in old calculations of the disk-wind
interaction. Our results may have implications for the variability
observed in flaring XRBs where a strong radio jet appears only
when the system transitions from the hard to the soft high state
(Kylafis et al.~2012).

Similar complex variability is also observed in the numerical
simulations of Parfrey et al.~(2015) which are similar in concept
to the ones presented in this work, although they solve only for
the magnetospheric field. In the latter, the magnetic flux loops
that inflate and open up are sustained by the disk turbulence as a
boundary condition, and this is the reason highly variable
reconnecting current sheets appear everywhere throughout their
simulations. Our physical picture differs in that we do account
for the origin of a large scale magnetic flux of a well defined
polarity that threads the disk and the inner compact object. In
our case, only one reconnecting current sheet develops at the
interface between outgoing and returning magnetic field lines.

Obviously, more investigation is required. Our work will be
extended soon with general relativistic simulations that take into
account the full distorted radiation field generated by the
innermost hot accretion flow around the central spinning black
hole (Koutsantoniou \& Contopoulos, in preparation).

\acknowledgements

This work was supported by the General Secretariat for Research
and Technology of Greece and the European Social Fund in the
framework of Action `Excellence'.

\begin{deluxetable}{lcccl}
\tablecolumns{5} \tablewidth{0pc} \tablecaption{Timescales to
reach equipartition} \tablehead{ \colhead{$M$} & \colhead{$B_{\rm
acc}$} & \colhead{$t_{\rm acc}$}
& \colhead{$B_{\rm eq}$} & \colhead{$t_{\rm eq}$}}\\ \\
 \startdata
1    &  0.1  &  $10^{-3}$ &  $10^8$ &  {\rm one week}    \\
10    &  0.01  &  $10^{-2}$ &   $10^7$ & {\rm a few months} \\
$10^9$  &  $10^{-9}$ & $10^6$ &  $10^3$ & {\rm several Gyears}
\enddata
\tablecomments{$M$ (in solar masses): the mass of the central
black hole; $B_{\rm acc}$ (in Gauss): the magnetic field
accumulated inside the ISCO in one accretion time $t_{\rm acc}$
(in seconds); $B_{\rm eq}$ (in Gauss): the nominal equipartition
magnetic field; $t_{\rm eq}$: the time needed for the Cosmic
Battery to build $B_{\rm eq}$.}
\end{deluxetable}

{}

\end{document}